\documentclass[10pt]{article}

\usepackage[totalheight = 23cm, totalwidth = 17cm]{geometry}
\usepackage{amssymb,amsmath,amsfonts,amsbsy,graphicx,bm}
\usepackage{ulem}
\usepackage{color}

\newcommand{\calN}{\mathcal N}

\begin{document}

\begin{titlepage}

\rightline{\footnotesize{APCTP-Pre2017-009}}
\vspace{-0.2cm}

\begin{center}

\vskip 1.0 cm

{\Huge \bf
Exact non-linear equations \\ for cosmological perturbations
}

\vskip 1.0cm

{\large
Jinn-Ouk Gong$^{a,b}$,
Jai-chan Hwang$^{c}$,
Hyerim Noh$^{d}$,
David Chan Lon Wu$^{e}$
and Jaiyul Yoo$^{e,f}$
}

\vskip 0.5cm

{\it
$^{a}$Asia Pacific Center for Theoretical Physics, Pohang 37673, Korea
\\
$^{b}$Department of Physics, Postech, Pohang 37673, Korea
\\
$^{c}$Department of Astronomy and Atmospheric Sciences,
\\
Kyungpook National University, Daegu 41566, Korea
\\
$^{d}$Korea Astronomy and Space Science Institute, Daejeon 34055, Korea
\\
$^{e}$Center for Theoretical Astrophysics and Cosmology, Institute for Computational Science,
\\
Universit\"at Z\"urich, CH-8057 Z\"urich, Switzerland
\\
$^{f}$Physics Institute, Universit\"at Z\"urich, CH-8057 Z\"urich, Switzerland
}

\vskip 1.2cm

\end{center}

\begin{abstract}

We present a complete set of exact and fully non-linear equations describing all three types of cosmological perturbations -- scalar, vector and tensor perturbations. We derive the equations in a thoroughly gauge-ready manner, so that any spatial and temporal gauge conditions can be employed. The equations are completely general without any physical restriction except that we assume a flat homogeneous and isotropic universe as a background. We also comment briefly on the application of our formulation to the non-expanding Minkowski background.

\end{abstract}

\end{titlepage}

\newpage

\section{Introduction}

%linear perturbation theory works good
%need to explore non-linear & gr perturbation theory
%previous studies were not complete
%we provide complete equations

The theory of cosmological perturbations is based on the cosmological principle, viz. a homogeneous and isotropic universe on very large scales~\cite{Friedmann}. Namely, given an expanding homogeneous and isotropic background, we introduce small deviations from this geometric background as well as those in the corresponding matter sector, and study the evolution of these perturbations. First pioneered by Lifshitz in 1940s~\cite{Lifshitz:1945du}, the linear cosmological perturbation theory culminated in the astonishing agreement with the observations on the angular power spectrum of the temperature fluctuations and polarizations in the cosmic microwave background (CMB), e.g. see~\cite{Ade:2015xua}.

Furthermore, the recent advances in cosmological observations demand theoretical endevour to understand non-linear regime of cosmological perturbations. One reason is because even for the CMB observations the technical development has become mature enough to search for small but non-zero signature of the primordial non-Gaussianity, e.g.~\cite{Ade:2015ava}. Also, the large-scale distribution of galaxies, which is one of the prime observational probes to understand the physics of the universe, involves non-linear evolution to form gravitationally bound system like clusters of galaxies. This becomes obvious on small scales where the prediction of the linear theory deviates from the matter power spectrum from N-body simulations (see e.g.~\cite{Jain:1993jh}). Especially, as the planned galaxy surveys such as DESI~\cite{desiweb}, LSST~\cite{lsstweb} and Euclid~\cite{euclidweb} are to probe very large volume accessing the scales comparable to the horizon, it is called for that we describe the evolution of non-linear cosmological perturbations in fully general relativistic context.

With general relativity being a gauge theory, that is, we are allowed to adopt any coordinate system to describe the physics, it is essential to clarify which gauge we are using to formulate the evolution of cosmological perturbations and to interpret the data from surveys. A systematic approach to make use of physically motivated gauge choices with gauge-invariant perturbation variables was made in~\cite{Bardeen:1980kt}, and it was subsequently argued~\cite{bardeen1988} that one should make use of various gauges to investigate the problem most conveniently. This approach of keeping the advantage of gauge-ready formulation has been extended to second-order perturbations in~\cite{Noh:2004bc,Hwang:2007ni}, with the solutions for the scalar perturbations found in~\cite{Hwang1994,Hwang:2012bi,Yoo:2015uxa}. In particular, in~\cite{Hwang:2012aa} the perturbation equations were presented valid for {all} order in perturbations in a gauge-ready form. But~\cite{Hwang:2012aa} was incomplete in the sense that the tensor components of cosmological perturbations were not included and we miss the critical information on them, and that from the beginning a specific spatial gauge condition was chosen which leaves no gauge degrees of freedom [see \eqref{eq:spatialgauge}].

In this article, we provide exact and fully non-linear relativistic description of cosmological perturbation theory by including \textit{all} three types of perturbations without choosing any spatial and temporal gauge conditions. Our exact equations do not assume that cosmological perturbations are small and can be extended to arbitrary higher order in perturbations, so fully non-linear cosmological perturbations can be easily studied. Thus our purpose in this article is to provide the general and necessary building blocks for studying perturbations analytically and/or numerically in highly non-linear regime in cosmology (or astrophysics: see Section~\ref{sec:discussion}). The outline of this article is as follows. In Section~\ref{sec:inverse-h}, we first derive the exact inverse metric including tensor perturbations, which is an essential step to derive the exact perturbation equations for cosmological perturbations in Section~\ref{sec:nleqs}. We summarize the implications of our results in Section~\ref{sec:discussion}. Throughout the article, we set $c=1$.

\section{Derivation of exact inverse metric}
\label{sec:inverse-h}

The Arnowitt-Deser-Misner (ADM) formulation~\cite{Arnowitt:1962hi} is based on splitting the space-time into the spatial and temporal parts using the unit normal four-vector field $n_\mu$ to constant-time three-hypersurfaces,
\begin{equation}
n_\mu = (-N,0)
\quad \text{and} \quad
n^\mu = \left( \frac{1}{N}, -\frac{N^i}{N} \right) \, ,
\end{equation}
where $N$ and $N^i$ are respectively the lapse and the shift [see \eqref{eq:ADMelement}] with $n^\mu n_\mu = -1$. The ADM formulation provides the exact non-linear equations of general relativity. However, its application to cosmology has to be dealt with order by order in perturbations, because no exact solution for the inverse spatial metric has been ever derived before. Here we derive its exact expression, and in the following section present the exact equations for cosmological perturbations.

The line element can be written as
\begin{equation}
\label{eq:ADMelement}
ds^2 = -N^2(dx^0)^2 + h_{ij} \left( N^idx^0 + dx^i \right) \left( N^jdx^0 + dx^j \right) \, ,
\end{equation}
so that each component of the metric is exactly found as
\begin{equation}
\label{eq:ADMmetric}
\begin{split}
& g_{00} = -N^2 + N^iN_i \, , \quad g_{0i} = N_i \, , \quad g_{ij} = h_{ij} \, ,
\\
& g^{00} = -\frac{1}{N^2} \, , \quad g^{0i} = \frac{N^i}{N^2} \, , \quad g^{ij} = h^{ij} - \frac{N^iN^j}{N^2} \, ,
\end{split}
\end{equation}
with the spatial indices of the ADM variables being raised and lowered by the spatial metric $h_{ij}$, e.g. $N^i = h^{ij}N_j$.

Our metric convention of a flat Friedmann universe including fully non-linear perturbations is
\begin{equation}
\label{eq:metric-pert}
ds^2 = -a^2(1+2\alpha)d\eta^2 - 2a^2B_i d\eta dx^i
+ a^2 \left[ (1+2\varphi)\delta_{ij} + 2\gamma_{,ij} + 2C^{(v)}_{(i,j)} + 2C_{ij} \right] dx^idx^j \, ,
\end{equation}
where $dx^0 = d\eta \equiv dt/a$ is the conformal time and $a(\eta)$ is the cosmic scale factor. The superscript $(v)$ denotes transverse vector, and the pure tensor perturbation $C_{ij}$ is transverse and traceless (TT). The indices of the perturbation variables in \eqref{eq:metric-pert} are raised and lowered by $\delta_{ij}$ as the metric tensor. With $B_i$ being decomposed into the gradient of a scalar and a transverse vector as $B_i = \beta_{,i} + B_i^{(v)}$, the shear $\chi_i$ is written as
\begin{equation}
\label{eq:shear}
\chi_i \equiv a \Big( B_i + a\dot\gamma_{,i} + a\dot{C}_i^{(v)} \Big)
= a \big( \beta + a\dot\gamma \big)_{,i} + a \Big( B_i^{(v)} + a\dot{C}_i^{(v)} \Big)
\equiv \chi_{,i} + \chi_i^{(v)} \, .
\end{equation}
The perturbation variables are general functions of space and time, and we do not assume these to be small in amplitudes. As can be read from \eqref{eq:ADMmetric}, an essential step to write the fully non-linear, exact equations is to find the exact inverse spatial metric $h^{ij}$. Comparing with \eqref{eq:ADMmetric}, the exact inverse metric $g^{\mu\nu}$ then follows automatically.

For a general invertible $3\times3$ matrix $M_{ij}$ the inverse matrix can be written as, explicitly writing the cofactor,
\begin{equation}
\left( M^{-1} \right)_{ij} = \frac{\epsilon_{ilm}\epsilon_{jpq}M_{pl}M_{qm}}{2\det\left[ M_{ij} \right]}
\quad \text{with} \quad
\det\left[ M_{ij} \right] = \frac{1}{3!} \epsilon_{ilm}\epsilon_{jpq} M_{pl}M_{qm}M_{ij} \, .
\end{equation}
Then, with $Y_{ij} \equiv \gamma_{,ij} + C_{(i,j)}^{(v)} + C_{ij}$, we find
\begin{equation}
\begin{split}
\det\left[h_{ij}\right] & =
a^6 \Big[ (1+2\varphi+\Delta\gamma)^2 + (\Delta\gamma)^2 - 2Y^{kl}Y_{kl} \Big]
\Bigg\{ 1 + 2 \bigg[ \underbrace{ \varphi + \frac{2}{3} \frac{(\Delta\gamma)^3 - 3(\Delta\gamma)Y^{kl}Y_{kl} + 2Y_{kl}Y^k{}_mY^{lm}}
{(1+2\varphi+\Delta\gamma)^2 + (\Delta\gamma)^2 - 2Y^{pq}Y_{pq}} }_{\equiv \widehat\varphi} \bigg] \Bigg\} \, ,
\\
\epsilon^i{}_{lm}\epsilon^j{}_{pq}h_{pl}h_{qm} & =
2a^4 \Big[ (1+2\varphi+\Delta\gamma)^2 + (\Delta\gamma)^2 - 2Y^{kl}Y_{kl} \Big]
\Bigg[ \delta^{ij} \underbrace{ - 2 \frac{(1+2\varphi+2\Delta\gamma)Y^{ij} - 2Y^{ik}Y^j{}_k}
{(1+2\varphi+\Delta\gamma)^2 + (\Delta\gamma)^2 - 2Y^{lm}Y_{lm}} }_{\equiv H^{ij}} \Bigg] \, ,
\end{split}
\end{equation}
so that the exact inverse of the spatial metric is
\begin{equation}
\label{eq:inversehij}
h^{ij} = \frac{\delta^{ij} + H^{ij}}{a^2(1+2\widehat\varphi)} \, .
\end{equation}
The indices of $Y_{ij}$ and $H^{ij}$ are raised and lowered by $\delta_{ij}$. One can explicitly check that indeed $h^{ik}h_{kj} = \delta^i{}_j$, so \eqref{eq:inversehij} is the exact inverse spatial metric.

With \eqref{eq:inversehij}, by comparing with \eqref{eq:ADMmetric} we have
\begin{equation}
\begin{split}
N & = a \sqrt{1 + 2\alpha + \frac{\delta^{ij}+H^{ij}}{a^2(1+2\widehat\varphi)} \chi_i\chi_j} \equiv a\calN \, ,
\\
N_i & = -a\chi_i \, ,
\\
N^i & = -\frac{\delta^{ij}+H^{ij}}{1+2\widehat\varphi} \frac{\chi_j}{a} \, ,
\end{split}
\end{equation}
so that each component of the inverse metric is exactly given by
\begin{equation}
\begin{split}
g^{00} & = -\frac{1}{a^2\calN^2} \, ,
\\
g^{0i} & = -\frac{\delta^{ij}+H^{ij}}{a^2\calN^2(1+2\widehat\varphi)} \frac{\chi_j}{a} \, ,
\\
g^{ij} & = \frac{1}{a^2(1+2\widehat\varphi)} \left[ \delta^{ij} + H^{ij}
- \frac{(\delta^{ik}+H^{ik})(\delta^{jl}+H^{jl})}{a^2\calN^2(1+2\widehat\varphi)} \chi_k\chi_l \right] \, .
\end{split}
\end{equation}
Having obtained the exact inverse metric, now we can proceed to find the fully non-linear and exact perturbation equations.

\section{Exact non-linear equations}
\label{sec:nleqs}

Here we present the exact non-linear equations for cosmological perturbations. After a brief introduction of the ADM equations in Section~\ref{subsec:ADM}, we present the ADM quantities in terms of cosmological perturbations in Section~\ref{subsec:quantities}. A complete set of exact equations of cosmological perturbation is presented in Section~\ref{subsec:eqs}.

\subsection{ADM equations}
\label{subsec:ADM}

In the ADM formulation, the dynamics of the space-time is described by the spatial metric $h_{ij}$ through the curvature variables of the spatial hypersurfaces. Along with the matter contents residing in the space-time, the geometric equations for the curvature variables constitute a complete set of the equations of motion. The extrinsic curvature $K_{ij}$ is introduced as
\begin{equation}
K_{ij} \equiv \frac{1}{2N} \left( N_{i:j} + N_{j:i} - h_{ij,0} \right) \, ,
\end{equation}
where a colon denotes a covariant derivative based on $h_{ij}$. The trace $K$ and trace-free part $\overline{K}_{ij}$ are respectively
\begin{equation}
K \equiv h^{ij}K_{ij}
\quad \text{and} \quad
\overline{K}_{ij} \equiv K_{ij} - \frac{1}{3}h_{ij}K \, .
\end{equation}
The connection based on $h_{ij}$ as the metric is written as $\Gamma^i_{jk} \equiv h^{il}/2 \left( h_{jl,k}+h_{kl,j}-h_{jk,l} \right)$, from which the intrinsic curvature tensor and scalar are given by
\begin{equation}
\begin{split}
R^i{}_{jkl} & \equiv \Gamma^i_{jl,k} - \Gamma^i_{jk,l} + \Gamma^m_{jl}\Gamma^i_{km} - \Gamma^m_{jk}\Gamma^i_{lm} \, ,
\\
R_{ij} & \equiv R^k{}_{ikj} \, ,
\\
R & \equiv h^{ij}R_{ij} \, .
\end{split}
\end{equation}
The trace-free part $\overline{R}_{ij}$ is written in the same way as $\overline{K}_{ij}$, i.e. $\overline{R}_{ij} = R_{ij} - h_{ij}R/3$.

The energy-momentum tensor is expanded in terms of the fluid quantities as
\begin{equation}
T_{\mu\nu} = (\mu+p)u_\mu u_\nu + pg_{\mu\nu} + \pi_{\mu\nu} \, ,
\end{equation}
with the unit fluid four-velocity satisfying $u^\mu u_\mu = -1$ and the anisotropic stress satisfying $\pi^\mu{}_\mu = 0$ and $\pi_{\mu\nu}u^\mu = 0$. The ADM fluid quantities are defined in terms of the energy-momentum tensor and the normal vector as
\begin{equation}
\begin{split}
E & \equiv n_\mu n_\nu T^{\mu\nu} \, ,
\\
J_i & \equiv -n_\mu T^\mu{}_i \, ,
\\
S_{ij} & \equiv T_{ij} \, ,
\end{split}
\end{equation}
where the indices of the momentum density $J_i$ and the spatial energy-momentum tensor $S_{ij}$ are raised and lowered by $h_{ij}$. We can also construct in the same way as for $K_{ij}$ and $R_{ij}$ the trace and trace-free part of $S_{ij}$, denoted by $S$ and $\overline{S}_{ij}$ respectively.

A complete set of the six ADM equations can be written exactly in terms of these ADM geometric and fluid quantities as follows~\cite{Bardeen:1980kt,Noh:2004bc}:
\begin{enumerate}

\item Energy constraint: This equation relates the energy density of the fluid to the geometry, i.e. curvature,
\begin{equation}
R = \overline{K}^i{}_j\overline{K}^j{}_i - \frac{2}{3}K^2 + 16\pi GE + 2\Lambda \, ,
\label{eq:ADM-E-const}
\end{equation}
where $\Lambda$ is a cosmological constant.

\item Momentum constraint: This equation gives the relation between the momentum density and the curvature,
\begin{equation}
\overline{K}^j{}_{i:j} - \frac{2}{3}K_{,i} = 8\pi GJ_i \, .
\end{equation}

\item Trace evolution: This gives the trace part of the equation of motion for the spatial metric $h_{ij}$. The evolution of the extrinsic curvature $K$ is described by this equation,
\begin{equation}
\frac{K_{,0}}{N} - \frac{K_{,i}N^i}{N} + \frac{N^{:i}{}_{:i}}{N} - \overline{K}^i{}_j\overline{K}^j{}_i
- \frac{1}{3}K^2 - 4\pi G (E+S) + \Lambda = 0 \, .
\end{equation}

\item Trace-free evolution: This is the traceless part of the equation of motion of $h_{ij}$, in terms of the traceless part of the extrinsic curvature $\overline{K}_{ij}$. As the name stands, this equation identically vanishes for $i=j$,
\begin{equation}
\frac{\overline{K}^i{}_{j,0}}{N} - \frac{\overline{K}^i{}_{j:k}N^k}{N} + \frac{\overline{K}^k{}_jN^i{}_{:k}}{N}
- \frac{\overline{K}^i{}_kN^k{}_{:j}}{N} = K\overline{K}^i{}_j
- \frac{1}{N} \left( N^{:i}{}_{:j} - \frac{\delta^i{}_j}{3}N^{:k}{}_{:k} \right) + \overline{R}^i{}_j
- 8\pi G \overline{S}^i{}_j \, .
\end{equation}

\item Energy conservation: This equation corresponds to the time component of the energy-momentum tensor conservation equation and tells us how the energy density evolves,
\begin{equation}
\frac{E_{,0}}{N} - \frac{E_{,i}N^i}{N} - K \left( E + \frac{1}{3}S \right)
- \overline{S}^i{}_j\overline{K}^j{}_i + \frac{\left(N^2J^i\right)_{:i}}{N^2} = 0 \, .
\end{equation}

\item Momentum conservation: This spatial component of the conservation equation gives the equation of motion of the three-velocity of the fluid,
\begin{equation}
\frac{J_{i,0}}{N} - \frac{J_{i:j}N^j}{N} - \frac{J_jN^j{}_{:i}}{N} - KJ_i
+ \frac{EN_{,i}}{N} + S^j{}_{i:j} + \frac{S^j{}_iN_{,j}}{N} = 0 \, .
\label{eq:ADM-Mom-cons}
\end{equation}

\end{enumerate}
These equations are exact and fully non-linear. To investigate, however, the evolution of cosmological perturbations we need to express these equations in terms of the standard variables for cosmological perturbations presented in the previous section.

\subsection{ADM quantities}
\label{subsec:quantities}

Here we present the ADM  quantities in terms of the perturbation variables.

The normal four-vector is
\begin{equation}
\begin{split}
n_\mu & = (-a\calN,0) \, ,
\\
n^\mu & = \left( \frac{1}{a\calN}, \frac{\delta^{ij}+H^{ij}}{a\calN(1+2\widehat\varphi)}\frac{\chi_j}{a} \right) \, .
\end{split}
\end{equation}
The intrinsic three-curvature tensor and scalar are
\begin{align}
R_{ij} & = \left[ \frac{\delta^{kl}+H^{kl}}{1+2\widehat\varphi}
\left( \varphi_{,i}\delta_{jl} + \varphi_{,j}\delta_{il} - \varphi_{,l}\delta_{ij}
+ Y_{jl,i} + Y_{il,j} - Y_{ij,l} \right) \right]_{,k}
- \left[ \frac{\delta^{kl}+H^{kl}}{1+2\widehat\varphi} \big( \varphi_{,i}\delta_{kl} + Y_{kl,i} \big) \right]_{,j}
\nonumber\\
& \quad + \frac{\delta^{lm}+H^{lm}}{(1+2\widehat\varphi)^2} \left(
\varphi_{,i}\delta_{jm} + \varphi_{,j}\delta_{im} - \varphi_{,m}\delta_{ij}
+ Y_{jm,i} + Y_{im,j} - Y_{ij,m} \right)
\big( \delta^{kn} + H^{kn} \big) \big( \varphi_{,l}\delta_{kn} + Y_{kn,l} \big)
\nonumber\\
& \quad - \frac{(\delta^{km}+H^{km})(\delta^{ln}+H^{ln})}{(1+2\widehat\varphi)^2}
\left( \varphi_{,i}\delta_{lm} + \varphi_{l}\delta_{im} - \varphi_{,m}\delta_{il}
+ Y_{lm,i} + Y_{im,l} - Y_{il,m} \right)
\nonumber\\
& \hspace{13em} \times \left( \varphi_{,j}\delta_{kn} + \varphi_{,k}\delta_{jn} - \varphi_{,n}\delta_{jk}
+ Y_{kn,j} + Y_{jn,k} - Y_{jk,n} \right) \, ,
\\
R & = \frac{\delta^{ij}+H^{ij}}{a^2(1+2\widehat\varphi)} R_{ij} \, ,
\\
\overline{R}^i{}_j & = \frac{\delta^{ik}+H^{ik}}{a^2(1+2\widehat\varphi)}R_{jk}
- \frac{\delta^i{}_j}{3} R \, .
\end{align}
The extrinsic three-curvature tensor and scalar are
\begin{align}
K_{ij} & = -\frac{a^2}{\calN} \left[ \left( H+ \dot\varphi + 2H\varphi \right)\delta_{ij} + \dot{Y}_{ij} + 2HY_{ij}
+ \frac{\chi_{(i,j)}}{a^2} - \frac{\delta^{kl}+H^{kl}}{a^2(1+2\widehat\varphi)} \chi_k
\left( \varphi_{,i}\delta_{jl} + \varphi_{,j}\delta_{il} - \varphi_{,l}\delta_{ij}
+ Y_{jl,i} + Y_{il,j} - Y_{ij,l} \right) \right] \, ,
\\
\label{eq:kappa}
K & = -3H + 3H \bigg( 1-\frac{1}{\calN} \bigg) - \frac{\delta^{ij}+H^{ij}}{\calN(1+2\widehat\varphi)}
\left[ \dot\varphi\delta_{ij} + \dot{Y}_{ij} + \frac{\chi_{i,j}}{a^2}
- \frac{\delta^{kl}+H^{kl}}{a^2(1+2\widehat\varphi)} \chi_k \left( 2\varphi_{,i}\delta_{jl} - \varphi_{,l}\delta_{ij}
+ 2Y_{il,j} - Y_{ij,l} \right) \right]
\nonumber\\
& \equiv -3H + \kappa \, ,
\\
\overline{K}^i{}_j & = - \frac{\delta^{ik}+H^{ik}}{\calN(1+2\widehat\varphi)} \left[ \dot\varphi\delta_{jk} + \dot{Y}_{jk}
+ \frac{\chi_{(j,k)}}{a^2} - \frac{\delta^{lm}+H^{lm}}{a^2(1+2\widehat\varphi)} \chi_m
\left( \varphi_{,j}\delta_{kl} + \varphi_{,k}\delta_{jl} - \varphi_{,l}\delta_{jk}
+ Y_{kl,j} + Y_{jl,k} - Y_{jk,l} \right) \right]
\nonumber\\
& \quad + \frac{\delta^i{}_j}{3} \frac{\delta^{kl}+H^{kl}}{\calN(1+2\widehat\varphi)}
\left[ \dot\varphi\delta_{kl} + \dot{Y}_{kl} + \frac{\chi_{k,l}}{a^2}
- \frac{\delta^{mn}+H^{mn}}{a^2(1+2\widehat\varphi)} \chi_m \left( 2\varphi_{,k}\delta_{ln} - \varphi_{,n}\delta_{kl}
+ 2Y_{kn,l} - Y_{kl,n} \right) \right] \, .
\end{align}

The fluid three-velocity $V^i$ measured by the Eulerian observer with normal four-vector $n_\mu$ is defined by~\cite{Banyuls:1997zz}
\begin{equation}
V^i \equiv \frac{h^{(n)}{}^i{}_\mu u^\mu}{-n_\mu u^\mu} = \frac{1}{N} \left( \frac{u^i}{u^0}+N^i \right) \, ,
\end{equation}
where $h^{(n)}_{\mu\nu} = g_{\mu\nu}+n_\mu n_\nu$ is the projection tensor normal to $n_\mu$. The index of $V^i$ is raised and lowered by $h_{ij}$. We introduce the perturbed fluid velocity as
\begin{equation}
V_i = h_{ij}V^j \equiv av_i \, ,
\end{equation}
whose index is raised and lowered by $\delta_{ij}$. Then each component of the fluid four-velocity is
\begin{equation}
\begin{split}
& u_i = a\gamma v_i \, ,
\quad
u_0 = -a\calN\gamma \left[ 1 + \frac{\delta^{ij}+H^{ij}}{a\calN(1+2\widehat\varphi)} \chi_iv_j \right] \, ,
\\
& u^i = \frac{\delta^{ij}+H^{ij}}{a(1+2\widehat\varphi)} \gamma \left( v_j + \frac{\chi_j}{a\calN} \right) \, ,
\quad
u^0 = \frac{\gamma}{a\calN} \, ,
\end{split}
\end{equation}
where we have introduced the Lorentz factor
\begin{equation}
\gamma \equiv -u^\mu n_\mu = \left( 1 - \frac{\delta^{ij}+H^{ij}}{1+2\widehat\varphi}v_iv_j \right)^{-1/2} \, .
\end{equation}

With the anisotropic stress introduced as
\begin{equation}
\begin{split}
\pi_{ij} & \equiv a^2\Pi_{ij} \, ,
\\
\pi^0{}_i & = a\frac{\delta^{jk}+H^{jk}}{\calN(1+2\widehat\varphi)}v_k\Pi_{ij} \, ,
\\
\pi^0{}_0 & = -\frac{(\delta^{ik}+H^{ik})(\delta^{jl}+H^{jl})}{(1+2\widehat\varphi)^2}
v_k \left( v_l + \frac{\chi_l}{a\calN} \right) \Pi_{ij} \, ,
\\
\pi^i{}_j & = \frac{\delta^{ik}+H^{ik}}{1+2\widehat\varphi}\Pi_{jk}
+ \frac{(\delta^{ik}+H^{ik})(\delta^{lm}+H^{lm})}{(1+2\widehat\varphi)^2}
\frac{\chi_k}{a\calN} v_m \Pi_{jl} \, ,
\end{split}
\end{equation}
the energy-momentum tensor becomes
\begin{align}
T^0{}_0 & = -\mu - (\mu+p) \left( \gamma^2-1
+ \frac{\delta^{ij}+H^{ij}}{1+2\widehat\varphi} \frac{\chi_i}{a\calN} \gamma^2v_j \right)
- \frac{(\delta^{ij}+H^{ij})(\delta^{kl}+H^{kl})}{(1+2\widehat\varphi)^2} v_j
\left( v_l + \frac{\chi_l}{a\calN} \right) \Pi_{ik} \, ,
\\
T^0{}_i & = (\mu+p)\gamma^2v_i + \frac{\delta^{jk}+H^{jk}}{\calN(1+2\widehat\varphi)} v_k\Pi_{ij} \, ,
\\
T_{ij} & = a^2 \Big[ (1+2\varphi)p\delta_{ij} + (\mu+p)\gamma^2v_iv_j + 2pY_{ij} + \Pi_{ij} \Big] \, .
\end{align}
The ADM fluid quantities can be written as
\begin{align}
E & = \mu + (\mu+p) \left( \gamma^2-1 \right)
+ \frac{(\delta^{ij}+H^{ij})(\delta^{kl}+H^{kl})}{(1+2\widehat\varphi)^2} v_jv_l\Pi_{ik} \, ,
\\
J_i & = a(\mu+p)\gamma^2v_i + a\frac{\delta^{jk}+H^{jk}}{1+2\widehat\varphi}v_k\Pi_{ij} \, ,
\\
S^i{}_j & = p\delta^i{}_j + \frac{\delta^{ik}+H^{ik}}{1+2\widehat\varphi}
\left[ (\mu+p)\gamma^2v_jv_k + \Pi_{jk} \right] \, .
\end{align}

\subsection{Exact equations for cosmological perturbations}
\label{subsec:eqs}

Given the expressions for the ADM geometric and fluid quantities in terms of cosmological perturbations, now we can simply substitute them in the exact fully non-linear equations presented in Section~\ref{subsec:ADM}. With the perturbation in the scalar extrinsic curvature $\kappa$ being given in \eqref{eq:kappa}, a complete set of equations follows from the ADM equations \eqref{eq:ADM-E-const}-\eqref{eq:ADM-Mom-cons} respectively:
\begin{equation}
R = \overline{K}^i{}_j\overline{K}^j{}_i - \frac{2}{3} \left( 9H^2 - 6H\kappa + \kappa^2 \right)
+ 16\pi G \left[ \mu + (\mu+p) \left( \gamma^2-1 \right)
+ \frac{(\delta^{ij}+H^{ij})(\delta^{kl}+H^{kl})}{(1+2\widehat\varphi)^2} v_jv_l \Pi_{ik} \right]
+ 2\Lambda \, ,
\end{equation}
\begin{align}
& \overline{K}^j{}_{i,j} + \frac{\delta^{kl} + H^{kl}}{1+2\widehat\varphi}
\big( \varphi_{,j}\delta_{kl} + Y_{kl,j} \big) \overline{K}^j{}_i
- \frac{\delta^{kl}+H^{kl}}{1+2\widehat\varphi} \overline{K}^j{}_k
\left( \varphi_{,j}\delta_{il} + \varphi_{,i}\delta_{jl} - \varphi_{,j}\delta_{ij}
+ Y_{il,j} + Y_{jl,i} - Y_{ij,l} \right) - \frac{2}{3}\kappa_{,i}
\nonumber\\
& = 8\pi G \left[ a(\mu+p)\gamma^2 v_i
+ a\frac{\delta^{jk}+H^{jk}}{1+2\widehat\varphi} v_k \Pi_{ij} \right] \, ,
\end{align}
\begin{align}
& -\frac{3}{\calN}\dot{H} + \frac{1}{\calN}
\left[ \dot\kappa + \frac{\delta^{ij}+H^{ij}}{a^2(1+2\widehat\varphi)} \chi_j\kappa_{,i} \right]
- \overline{K}^i{}_j\overline{K}^j{}_i + \left( -3H^2 + 2H\kappa - \frac{\kappa^2}{3} \right)
\nonumber\\
& + \frac{\delta^{ij}+H^{ij}}{a^2\calN(1+2\widehat\varphi)}
\left[ \calN_{,ij} - \frac{\delta^{kl}+H^{kl}}{1+2\widehat\varphi}
\left( \varphi_{,i}\delta_{jl} + \varphi_{,j}\delta_{il} - \varphi_{,l}\delta_{ij}
+ Y_{il,j} + Y_{jl,i} - Y_{ij,l} \right) \calN_{,k} \right]
\nonumber\\
& - 4\pi G \left\{ \mu + 3p + (\mu+p) \left( \gamma^2-1 \right)
+ \frac{\delta^{ij}+H^{ij}}{1+2\widehat\varphi}
\left[ (\mu+p)\gamma^2 v_iv_j + \Pi_{ij}
+ \frac{\delta^{kl}+H^{kl}}{1+2\widehat\varphi} v_jv_l \Pi_{ik} \right] \right\}
+ \Lambda = 0 \, ,
\end{align}
\begin{align}
& \left[ \frac{1}{\calN}\frac{\partial}{\partial t}
+ \frac{\delta^{kl}+H^{kl}}{a^2\calN(1+2\widehat\varphi)} \chi_l \frac{\partial}{\partial x^k}
+ (3H-\kappa) \right] \overline{K}^i{}_j
- \frac{1}{a^2\calN} \left( \frac{\delta^{il}+H^{il}}{1+2\widehat\varphi}\chi_l \right)_{,k} \overline{K}^k{}_j
+ \frac{1}{a^2\calN} \left( \frac{\delta^{kl}+H^{kl}}{1+2\widehat\varphi}\chi_l \right)_{,j} \overline{K}^i{}_k
\nonumber\\
= & -\frac{\delta^{ik}+H^{ik}}{a^2\calN(1+2\widehat\varphi)}
\left[ \calN_{,jk} - \frac{\delta^{lm}+H^{lm}}{1+2\widehat\varphi}
\left( \varphi_{,j}\delta_{km} + \varphi_{,k}\delta_{jm} - \varphi_{,m}\delta_{jk}
+ Y_{jm,k} + Y_{km,j} - Y_{jk,m} \right) \calN_{,l} \right]
\nonumber\\
& + \frac{\delta^i{}_j}{3} \frac{\delta^{kl}+H^{kl}}{a^2\calN(1+2\widehat\varphi)}
\left[ \calN_{,kl} - \frac{\delta^{mn}+H^{mn}}{1+2\widehat\varphi}
\left( \varphi_{,k}\delta_{ln} + \varphi_{,l}\delta_{kn} - \varphi_{,n}\delta_{kl}
+ Y_{ln,k} + Y_{kn,l} - Y_{kl,n} \right) \calN_{,m} \right]
\nonumber\\
& + \overline{R}^i{}_j - 8\pi G
\left\{ \frac{\delta^{ik}+H^{ik}}{1+2\widehat\varphi} \left[ (\mu+p)\gamma^2 v_jv_k + \Pi_{jk} \right]
- \frac{\delta^i{}_j}{3} \frac{\delta^{kl}+H^{kl}}{1+2\widehat\varphi}
\left[ (\mu+p)\gamma^2 v_kv_l + \Pi_{kl} \right] \right\} \, ,
\end{align}
\begin{align}
& \frac{1}{\calN} \left[ \frac{\partial}{\partial t} +
\frac{\delta^{ij}+H^{ij}}{a^2(1+2\widehat\varphi)}\chi_j\frac{\partial}{\partial x^i} \right]
\left[ \mu + (\mu+p) \left( \gamma^2-1 \right) +
\frac{(\delta^{kl}+H^{kl})(\delta^{mn}+H^{mn})}{(1+2\widehat\varphi)^2} v_lv_n \Pi_{km} \right]
\nonumber\\
& + (3H-\kappa) \left\{ (\mu+p)\gamma^2 + \frac{\delta^{ij}+H^{ij}}{1+2\widehat\varphi}
\left[ \frac{1}{3}(\mu+p)\gamma^2 v_iv_j + \frac{1}{3}\Pi_{ij}
+ \frac{\delta^{kl}+H^{kl}}{1+2\widehat\varphi} v_jv_l \Pi_{ik} \right] \right\}
\nonumber\\
& - \left\{ \frac{\delta^{ik}+H^{ik}}{1+2\widehat\varphi} \left[ (\mu+p) \gamma^2 v_jv_k + \Pi_{jk} \right]
- \frac{\delta^i{}_j}{3} \frac{\delta^{kl}+H^{kl}}{1+2\widehat\varphi}
\left[ (\mu+p) \gamma^2 v_kv_l + \Pi_{kl} \right] \right\} \overline{K}^j{}_i
\nonumber\\
& + \frac{\delta^{ij}+H^{ij}}{a(1+2\widehat\varphi)} \bigg\{ \left[ (\mu+p) \gamma^2 v_i
+ \frac{\delta^{kl}+H^{kl}}{1+2\widehat\varphi} v_l \Pi_{ik} \right]_{,j}
+ 2 \left[ (\mu+p) \gamma^2 v_i + \frac{\delta^{kl}+H^{kl}}{1+2\widehat\varphi} v_l \Pi_{ik} \right]
\frac{\calN_{,j}}{\calN}
\nonumber\\
& \qquad - \frac{\delta^{kl}+H^{kl}}{1+2\widehat\varphi} \left( \varphi_{,i}\delta_{jl} + \varphi_{,j}\delta_{il}
- \varphi_{,l}\delta_{ij} + Y_{jl,i} + Y_{il,j} - Y_{ij,l} \right) \left[ (\mu+p) \gamma^2 v_k
+ \frac{\delta^{mn}+H^{mn}}{1+2\widehat\varphi} v_n \Pi_{im} \right] \bigg\} = 0 \, ,
\end{align}
\begin{align}
\label{eq:momentumconservation}
& \left[ \frac{1}{\calN} \frac{\partial}{\partial t} +
\frac{\delta^{jk}+H^{jk}}{a^2\calN(1+2\widehat\varphi)} \chi_k \frac{\partial}{\partial x^j}
+ (3H-\kappa) \right] \left\{ a \left[ (\mu+p)\gamma^2 v_i
+ \frac{\delta^{mn}+H^{mn}}{1+2\widehat\varphi} v_n \Pi_{im} \right] \right\}
\nonumber\\
& + \frac{\delta^{jk}+H^{jk}}{a\calN(1+2\widehat\varphi)} \bigg\{ \chi_{k,i}
\left[ (\mu+p) \gamma^2 v_j + \frac{\delta^{mn}+H^{mn}}{1+2\widehat\varphi} v_n \Pi_{jm} \right]
\nonumber\\
& \qquad\qquad\qquad\quad - \frac{\delta^{lp}+H^{lp}}{1+2\widehat\varphi} \left( \varphi_{,i}\delta_{jp} + \varphi_{,j}\delta_{ip}
- \varphi_{,p}\delta_{ij} + Y_{jp,i} + Y_{ip,j} - Y_{ij,p} \right)
\nonumber\\
& \qquad\qquad\qquad\qquad \times \left[ \chi_k \left( (\mu+p) \gamma^2 v_l
+ \frac{\delta^{mn}+H^{mn}}{1+2\widehat\varphi} v_n \Pi_{lm} \right)
+ \chi_l \left( (\mu+p) \gamma^2 v_k
+ \frac{\delta^{mn}+H^{mn}}{1+2\widehat\varphi} v_n \Pi_{km} \right) \right] \bigg\}
\nonumber\\
& + \left[ \mu + (\mu+p) \left( \gamma^2-1 \right)
+ \frac{(\delta^{jm}+H^{jm})(\delta^{kl}+H^{kl})}{(1+2\widehat\varphi)^2} v_jv_l \Pi_{km} \right]
\frac{\calN_{,i}}{\calN}
\nonumber\\
& + \left[ \frac{\partial}{\partial x^j} + \frac{\delta^{kl} + H^{kl}}{1+2\widehat\varphi} \left( \varphi_{,j} \delta_{kl}
+ Y_{kl,j} \right) + \frac{\calN_{,j}}{\calN} \right]
\left\{ p\delta^j{}_i + \frac{\delta^{jm}+H^{jm}}{1+2\widehat\varphi} \left[ (\mu+p) \gamma^2 v_iv_m
+ \Pi_{im} \right] \right\}
\nonumber\\
& - \frac{\delta^{kl}+H^{kl}}{1+2\widehat\varphi} \left( \varphi_{,i}\delta_{kl} + \varphi_{,j}\delta_{il}
- \varphi_{,l}\delta_{ij} + Y_{jl,i} + Y_{il,j} - Y_{ij,l} \right) \left\{ p\delta^j{}_k
+ \frac{\delta^{jm}+H^{jm}}{1+2\widehat\varphi} \left[ (\mu+p) \gamma^2 v_kv_m + \Pi_{km} \right] \right\}
= 0 \, .
\end{align}
These exact and fully non-linear equations of cosmological perturbations are the main result of this article. Together with \eqref{eq:kappa} the above set of equations is complete in the sense that if we provide the equation of states, i.e. $p$ and $\Pi_{ij}$, after choosing spatial and temporal gauge conditions,  these equations can be solved exactly -- more practically, order by order perturbatively. We have checked that the non-linear equations in other literatures are successfully reproduced~\cite{othernonlinear}.

The spatial gauge condition\footnote{
Note that different spatial gauge conditions may have physical implications in the non-linear regime, such as Eulerian and Lagrangian observers in the comoving and comoving-synchronous gauge~\cite{Hwang:2014qfa,Yoo:2014vta}.
}
\begin{equation}
\label{eq:spatialgauge}
\gamma = 0 \quad \text{and} \quad C_i^{(v)} = 0
\end{equation}
completely eliminates the spatial gauge degree of freedom~\cite{Noh:2004bc} with the non-TT part of the spatial metric being isotropic, where we have $Y_{ij} = C_{ij}$. Together with a temporal gauge out of several choices (see below), \eqref{eq:spatialgauge} is unique in the sense that both the spatial and temporal gauge modes are completely fixed, thus each remaining variable has a unique gauge-invariant combination to all perturbation orders. Thus while not the only possible choice though, \eqref{eq:spatialgauge} can be regarded as physically transparent and convenient. A list of widely adopted temporal gauges is given in Table~\ref{table:gauge}, where we have decomposed, like the shear $\chi_i$ \eqref{eq:shear}, the velocity $v_i$ into the scalar gradient and transverse vector as $v_i = -v_{,i} + v_i^{(v)}$ and the energy density $\mu$ into the background and perturbation, $\mu = \mu_b+\delta\mu$. For example, under \eqref{eq:spatialgauge} using the comoving gauge in the matter-dominated universe where $p = 0$ and $\Pi_{ij}=0$ with vanishing vector perturbations, we have $v_i=0$ and the fluid quantities are greatly simplified to give
\begin{equation}
E = \mu
\quad \text{and} \quad
J_i = S_{ij} = 0 \, .
\end{equation}
This gives from the momentum conservation equation \eqref{eq:momentumconservation} $\calN$ is constant, which we may set to be 1 without losing generality. Thus the coordinate time precisely coincides with the proper time~\cite{Yoo:2014vta} so that we can up to second order, in the absence of the tensor perturbations, reproduce the Newtonian hydrodynamical equations, giving rise to the correspondence between the Newtonian and relativistic cosmologies~\cite{Noh:2004bc,Noh:2005hc}.

\begin{table}[htb!]
 \begin{center}
  \begin{tabular}{|c||c|}
   \hline
   Gauge & Condition
   \\
   \hline\hline
   Synchronous gauge & $\alpha = 0$
   \\
   \hline
   Comoving gauge & $v = 0$
   \\
   \hline
   Zero-shear gauge (Newtonian gauge) & $\chi = 0$
   \\
   \hline
   Uniform-expansion gauge & $\kappa = 0$
   \\
   \hline
   Uniform-curvature gauge (flat gauge) & $\varphi = 0$
   \\
   \hline
   Uniform-density gauge & $\delta\mu = 0$
   \\
   \hline
  \end{tabular}
 \end{center}
 \caption{A list of popular temporal gauges and the corresponding conditions.}
 \label{table:gauge}
\end{table}

Except for the synchronous gauge which fails to fix the temporal gauge degree of freedom even at linear order, the other gauge conditions completely fix the temporal (as well as the spatial) gauge degree of freedom to all perturbation orders. We may regard any variable under such gauge conditions as gauge-invariant in the sense that we can construct explicitly a unique corresponding gauge-invariant combination of the variables to all perturbation orders.

\section{Discussion}
\label{sec:discussion}

In this article, we have presented a complete set of exact fully non-linear equations for cosmological perturbations of all types, including the tensor perturbation without imposing any gauge condition. Considering the tensor perturbation and not imposing even the spatial gauge condition are the new features compared with previous studies in the cosmological context. As our equations now consider full Einstein's gravity without imposing any physical or coordinate condition the equations may be regarded as a new formulation of Einstein's gravity based on the flat spatial metric $\delta_{ij}$ in raising and lowering indices of the tensor variables appearing in the equations with arbitrary amplitudes. The equations can also be regarded as alternative presentation of the ADM equation with exact inverse of the intrinsic curvature, thus exact and suitable for perturbation expansion to nonlinear orders.

In the cosmological context, given the rapid development in experiments and observations in recent years,
more precise theoretical calculations are needed to meet the level of accuracy demanded by the current and future data set. In particular, higher-order statistics such as the bispectrum is routinely measured in current galaxy surveys and CMB experiments. Just to compute the tree-level bispectrum, we need the second-order equations for cosmological perturbations. While non-linear Newtonian perturbation theory is well developed, we need fully relativistic descriptions of cosmological observables (see e.g.~\cite{2nd-order-gr-pert}) to provide precise theoretical predictions such as the primordial non-Gaussianity and second-order relativistic effects. The goal in this article is to provide general exact relativistic equations for cosmological perturbations, so that any problem involving high non-linearity in cosmological perturbations can be solved as precisely as necessary.

We thus expect the formulation presented in this article has wide applications, not only in higher order perturbation in cosmology, but also in different problems, e.g. relativistic clustering of matter on small scales where the expansion of the universe can be neglected by setting $a = 1$ and $H = 0$, in which the background is reduced to the non-expanding Minkowski space-time. In the Minkowski background, among the gauge choices in Table~\ref{table:gauge} we can take the first four gauges, but the uniform-curvature and uniform-density gauges are not available~\cite{Hwang:2016kww}. Our exact fully non-linear equations describe precisely the evolution of perturbations with arbitrary large amplitude.

By including the tensor-type perturbation, now we can handle the higher order post-Newtonian (PN) approximation using our formulation. The radiation back-reaction is known to appear from 2.5 PN order~\cite{Chandrasekhar-Esposito-1970}. The higher order PN approximation treated in our formulation is an interesting topic, especially in a non-expanding background.

\subsection*{Acknowledgements}

We thank
Kyungjin Ahn and Chang Sub Shin
for discussions.
JG is grateful to the Center for Theoretical Astrophysics and Cosmology, Universit\"at Z\"urich for hospitality where this work was initiated.
JG acknowledges the support from the Korea Ministry of Education, Science and Technology, Gyeongsangbuk-Do and Pohang City for Independent Junior Research Groups at the Asia Pacific Center for Theoretical Physics. JG is also supported in part by a TJ Park Science Fellowship of POSCO TJ Park Foundation and the Basic Science Research Program through the National Research Foundation of Korea Research Grant (2016R1D1A1B03930408).
JH is supported by the Basic Science Research Program through the National Research Foundation of Korea (2016R1A2B4007964).
HN is supported by the National Resarch Foundation of Korea (2015R1A2A2A01002791).
DCLW and JY are supported by the Swiss National Science Foundation and a Consolidator Grant of the European Research Council (ERC-2015-CoG grant 680886).

\end{document}